\documentclass[usenatbib,usegraphicx,useAMS,a4]{mn2e}
\usepackage{amssymb}
\usepackage[totalwidth=490pt,totalheight=680pt]{geometry}
\newlength{\colwidth}
\newlength{\dantab}
\newlength{\proftab}
\def\fmins{\hbox{$.\mkern-4mu^{\prime}$}}

\begin{document}

\title {Decomposition of AGN host galaxy images}
\author [B. Kuhlbrodt, L. Wisotzki, K. Jahnke] 
   {B. Kuhlbrodt,$^{1,2}$ L. Wisotzki,$^{1,3}$ K. Jahnke$^{1,2}$\\
   $^1$Astrophysikalisches Institut Potsdam, An der Sternwarte 16, 
        14482 Potsdam, Germany\\
   $^2$Hamburger Sternwarte, Gojenbergsweg 112, 21029 Hamburg, Germany\\
   $^3$Institut f\"ur Physik, Potsdam University, Am Neuen Palais 10, 
        14469 Potsdam, Germany}

\maketitle

\begin{abstract}
  We describe an algorithm to decompose deep images of Active Galactic
  Nuclei into host galaxy and nuclear components. Currently supported
  are three galaxy models: A de-Vaucouleurs spheroidal, an exponential
  disc, and a two-component disc+bulge model.  Key features of the
  method are: (semi-)analytic representation of a possibly spatially
  variable point-spread function; full two-dimensional convolution of
  the model galaxy using gradient-controlled adaptive subpixelling;
  multiple iteration scheme. The code is computationally efficient and
  versatile for a wide range of applications. The quantitative
  performance is measured by analysing simulated imaging data.  We
  also present examples of the application of the method to small test
  samples of nearby Seyfert~1 galaxies and quasars at redshifts
  $z<0.35$. 
\end{abstract}

\begin{keywords}
  galaxies: active --  galaxies: photometry -- quasars: general
\end{keywords}

\section{Introduction}

The properties of black holes in galactic nuclei are probably 
closely linked to the global properties of the galaxies in which 
they reside. Fuelling these black holes leads to the
AGN and quasar phenomenon; investigating AGN host galaxies for various
degrees of AGN activity is therefore a necessary step to understand
the physical links, and the role of AGNs in galaxy evolution. Because
of the high luminosities of the central region -- being effectively a
point source in optical and near-infrared wavelengths -- which often
outshines the entire galaxy, quantitative study of quasar hosts is
fraught with technical difficulties. New instrumentation has made this
task somewhat more feasible. In particular HST with its high spatial 
resolution has contributed significantly to the study of quasar hosts 
both at low redshifts (\citealt{mclu99}; \citealt{mcle01}) 
and in the early universe \citep{kuku01,ridg01,lehn99}.  However,
ground-based imaging under excellent conditions will remain to be
competitive, especially with the new 8--10\,m class telescopes
\citep*[e.g.][]{falo01} using their large photon-collection area and
high resolution.

While the mere detection of QSO hosts often re\-quires no more than
elementary and intuitive methods such as azimuthal averaging and PSF
subtraction, such procedures have repeatedly been suspected of
producing quantitatively biased results (e.g., \citealt*{abra92};
\citealt{ravi01}). Quite certainly, they take insufficient advantage
of the full spatial image information content. In recent years, some
groups have started to develop two-dimensional model fitting codes
addressing these issues, with the goal to simultaneously decompose
deep QSO images into nuclear and host components in a more objective
and unbiased way (e.g., \citealt*{mclu00}; \citealt*{wada99};
\citealt*{scha96}). Ideally, such a method should provide the
flexibility to be used with a wide range of ground- and space-based
datasets, account for non-ideal detector properties, and require no
more than standard computing resources.

In this paper we describe our own approach to tackle this task. We
first outline some key features of the algorithm, and then discuss the
performance of our method as applied to simulated imaging data.
Finally we briefly present two samples of low- and
intermediate-redshift QSOs as examples of the method's usefulness.
The method is currently used extensively on various large datasets of
QSOs, achieving high data throughput for the modelling which is one of
the aims for our code.  We shall report in detail on these projects in
subsequent papers.

\section{The method}
\label{sec:meth}

\subsection{Overview}
\label{sec:over}

Optical and near-infrared images of quasars are always compounds of a
more or less extended host galaxy (which morphologically may be as
simple or as complicated as any `normal' galaxy), plus an embedded
point source. Analytic models of such configurations invariably
require several approximations and simplifications, which in our
approach can be summarised as follows:
\begin{itemize}
\item The overall surface brightness distribution of the host galaxy 
 can be described by smooth and azimuthally symmetric profile laws, 
 modified only to allow for a certain degree of ellipticity.
\item Host galaxy components and active nucleus (in the following: 
 `nucleus' or `AGN') are concentric.
\item The solid angle subtended by a given quasar+host is significantly
 smaller than the field of view.
\item The point-spread function (PSF) is either shift-invariant over the
 field of view, or else its spatial change can be described by low-order
 multivariate polynomials.
\end{itemize}
These assumptions are adequate for the type of distant AGN that we are
chiefly interested in, but some will break down for very nearby
galaxies with highly resolved structural features; such objects are
not our primary targets, and we do not consider their specialities in
the following.

The model-fitting process can be split up into several distinct tasks,
to be executed subsequently:
\begin{enumerate}
\item Construction of a variance frame quantifying individual pixel
weights, usually by applying Poisson statistics and standard
error propagation to object and background counts. This step 
includes the creation of an optional mask to exclude foreground stars, 
companion galaxies, cosmics, etc.
\item Identification of stars in the field to be used as PSF
references. As the PSF description is fully analytic, even stars
fainter than the quasar can yield useful constraints.
\item Determination of an analytic PSF model for the entire field of
view, accounting for spatial variability. An optional empirical
lookup table can complement this if required.
\item Establishing initial guess parameters for the AGN+host galaxy
model.
\item Computation of the actual multiparameter fit by minimising
 $\chi^2$ iteratively, including multiple restarts to avoid trapping
 in local minima.
\item Estimation of statistical uncertainties by running the
model-fitting code on dedicated simulations mimicking the actual data.
\end{enumerate}

We give details on each of these steps in the following sections.

\medskip The software was developed under the
ESO-MIDAS\footnote{http://www.eso.org/projects/esomidas/} environment
with all computing expensive tasks coded in C.
The code itself has not been developed for general public release,
but interested groups may receive a test version on a shared risk basis.

\subsection{PSF Modelling} \label{sec:psf}

\subsubsection{Strategy}

Knowledge of the point-spread function (PSF) is important in two
aspects of the decomposition. First, it is obviously needed to
describe the light distribution of the unresolved AGN itself. Any
mismatch here could lead to a misattribution of AGN light to the host
galaxy or vice versa. Second, for the typical objects of interest the
apparent host galaxy structure will strongly depend on the degree of
PSF blurring. This process needs somehow to be inverted in order to
determine the corresponding structural parameters. In extreme cases,
e.g.\ when even a marginal detection of a faint high-redshift host 
would be considered a success, accurate PSF control becomes the most 
important part of the entire analysis.

As long as the image formation process can be approximated by a
shift-invariant linear system, the straightforward and most frequently
adopted way of obtaining the PSF is to use the image of a bright star
in the field of view. However, even within this approximation using a
single star has some non-negligible drawbacks, mainly associated with
the problem of rebinning: Unless the PSF is
strongly oversampled, shifting an observed stellar image to a
different position invariably leads to image degradation and
consequently to AGN/PSF mismatch. Ironically, at a given spatial
sampling this effect is largest for a very narrow PSF, thus for the
best seeing. Furthermore, a single PSF star of sufficient brightness
to constrain also the low surface brightness wings of the PSF is not
always available, an effect which can render entire images effectively
useless. Finally, in a few cases even the only available PSF star
could be contaminated by a companion star or galaxy, which would
introduce severe artefacts into the analysis.

A simple averaging of stellar images to increase the S/N is
often prevented by the fact that several large-field
imagers, even modern ones, show spatial variations in the imaging
properties; in the above terminology, the system may still be linear
but not shift-invariant any more. Within the simple approach of
resampling PSF reference stars to the AGN position, there is only one
possible solution to this problem, namely limiting the allowable
distance AGN--PSF star to a minimum, and thereby often discarding the
brightest stars in the field.

To overcome this we adopted the alternative to describe the PSF
by an analytical expression, producing an essentially noise-free PSF
at any desired location with respect to the pixel grid.  An obvious
advantage of this approach is the fact that once a good analytical
description for a single star is found, averaging over several stars
is straightforward. In fact, since the main PSF parameters can be
measured confidently even at moderate S/N ratio, the number of
potential PSF stars usable is greatly increased, as now even stars
considerably fainter than the AGN can be used to provide constraints.

\begin{figure}
  \includegraphics[bb = 51 509 340 773,width=\colwidth,clip]{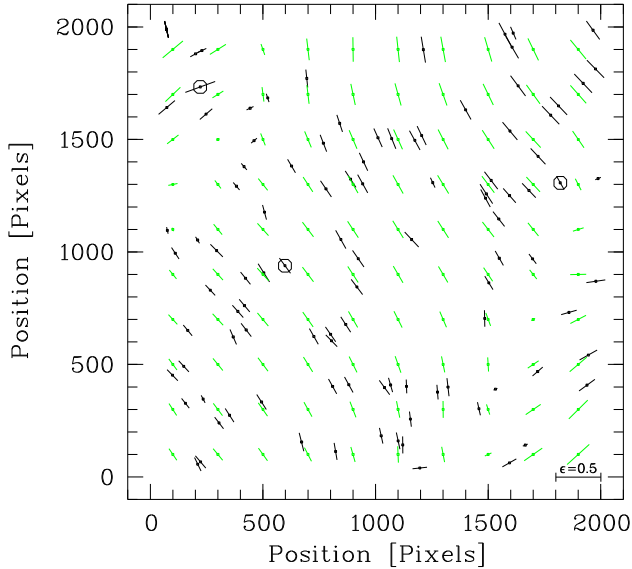}
  \caption{Visualisation of a spatially variable PSF. Each vector 
    corresponds to one star found in the image, its length given by
    the ellipticity and its orientation by the position angle of the
    major axis. Note the well-ordered pattern which makes analytic
    modelling straightforward, the resulting model grid is overlaid in
    light gray.  Circles mark the stars of
    Fig.~\protect{\ref{fig:psf}} in same order from left to right.
    Image size is 13\fmins{}3 across.}
  \label{fig:fish}
\end{figure}

\begin{figure}
 \includegraphics[width=\colwidth]{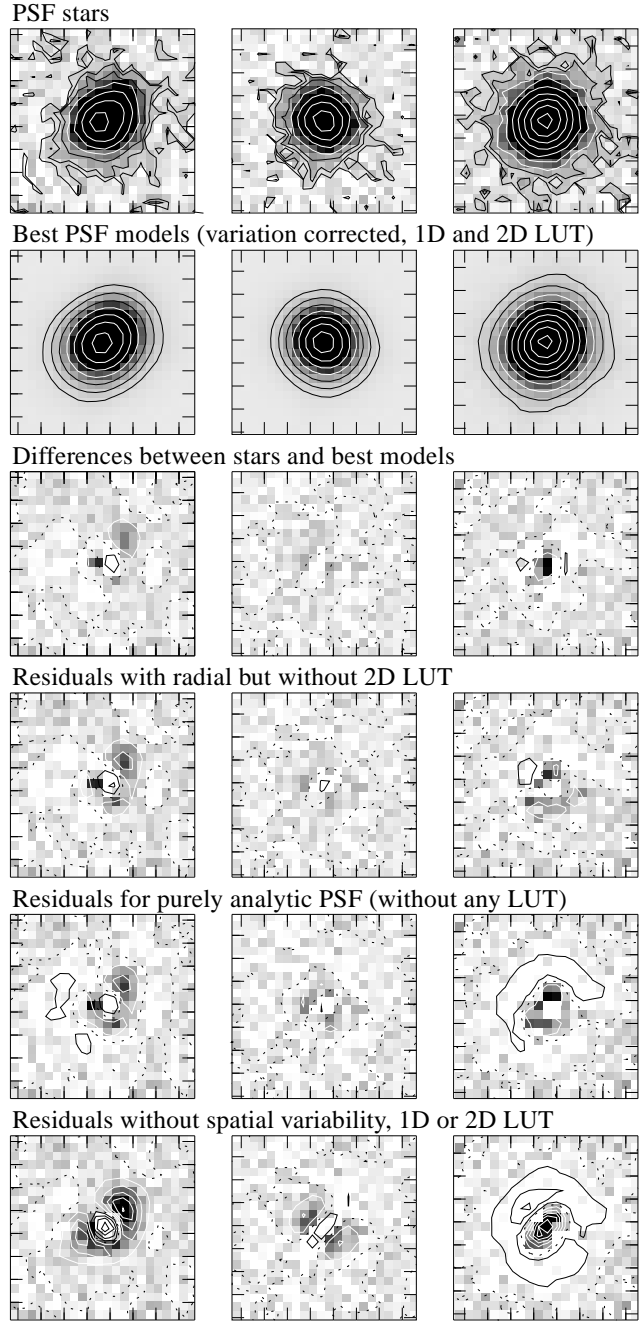}
  \caption[]{Modelling the PSF variations. Top row: Logarithmic contour plots and grayscale plots 
    of three example stars taken from different locations in the same
    image.  Slightly varying ellipticities can be traced even by eye.
    Second row: best models with modelling of spatial variation, one-
    and two-dimensional lookup table corrections. Third and subsequent
    rows: Residuals after subtracting decreasingly elaborate PSF
    models from each star.  Contours are linear and symmetric around
    zero (dotted line).  Coordinate tickmarks in all plots are
    0\farcs5 apart.}
   \label{fig:psf}
\end{figure}

In a straightforward generalisation of the analytic approach, the PSF
parameters can be described as spatially variable across the field. As
long as the variation model is adequate, all stars in the field can
still be used to trace and constrain the PSF. This is demonstrated in
Figs\ \ref{fig:fish} and \ref{fig:psf}, taken from our 1998 ESO data
documented below, but we have found similar effects with several other
instruments: While the ellipticities and orientations of point sources
in the field are obviously not constant, there is a discernible
variation pattern.  Once this pattern has been taken into account, the
overall PSF shape can be described by a well-constrained set of
parameters.

By choosing this approach, we consciously optimise our algorithm to
images with relatively simple PSF shapes, i.e.\ mainly ground-based
data without adaptive optics.  For instruments with a more complicated
PSF such as HST, a purely analytic point-symmetric PSF is clearly a gross
oversimplification.  However, departures from the symmetries assumed
in the analytic model can be accounted for by
applying a numerical lookup table correction (see Sect.\ \ref{sec:lut}
below).

\begin{table}
 \caption{Quality of different PSF models with or without correction of
    spatial variability,  one- and two dimensional LUTs. For the stars
    shown in Fig~\protect{\ref{fig:psf}} we tabulate the rms of residuals in 
    per cent of total stellar flux and the reduced $\chi^2$.
   \label{tab:psfrms}} 
 \begin{tabular}{ccc|rrrrrr}\hline
   \multicolumn{3}{c}{Correction} & \multicolumn{2}{c}{PSF 1}&\multicolumn{2}{c}{PSF 2}&\multicolumn{2}{c}{PSF 3}\\
   var.    & 1d.     & 2d.                  & $\Delta F$ & $\chi^2_\mathrm{r}$& $\Delta F$ & $\chi^2_\mathrm{r}$& $\Delta F$ & $\chi^2_\mathrm{r}$    \\\hline
   +       & +       & +                    & 0.066 & 2.2 & 0.059 & 1.6 & 0.029 & 1.7\\
   +       & +       & --                   & 0.095 & 2.6 & 0.060 & 1.7 & 0.034 & 1.8\\
   +       & --      & --                   & 0.101 & 2.9 & 0.077 & 1.8 & 0.039 & 2.3\\
   --      & --      & --                   & 0.166 & 6.3 & 0.096 & 2.2 & 0.060 & 3.7\\\hline
 \end{tabular}
\end{table}

\clearpage
\subsubsection{Analytic models}

To describe the radial PSF shape we have adopted Moffat's
(\citeyear{moff69}) PSF parameterisation, given in a modified form in
Eqn~\ref{eqn:moff1} below. We find that this profile provides a
reasonable fit to the PSF for several different datasets obtained in
both optical and NIR domains. Note that the Moffat parameter $\beta$,
which basically controls the kurtosis of the profile (larger $\beta$
implying a more peaked profile with weaker wings) has to be included
as a free parameter, as we often find best-fit $\beta$ values
significantly different from the canonical value of 2.5. Moffat's
original description has been reformulated to use $r_{1/2}$ as the
radius which encloses half the total flux:
\begin{eqnarray}
\label{eqn:moff1}
F_{\mathrm{PSF}} (r) &\:=\:& F_ {0,\mathrm{PSF}} \,
hscp des    \left[\,1+\frac{r^2}{r_{1/2}^2}\left(2^{1/\beta}-1\right)\right]^{-\beta}
\end{eqnarray}

Other analytic forms are conceivable, though the number of free
parameters should not be increased, as this requires to increase the
lower flux limit of acceptable stars which in consequence will
decrease the number of sampling points of the spatial PSF variation.
Instead, deviations between the analytic shape and the Moffat
prescription can be handled by a lookup table, described in the next
section.

The azimuthal PSF shape is assumed to be elliptical, thus requiring a
semimajor axis $a$, a semiminor axis $b$, and a position angle $\phi$
as additional parameters to specify the model.  We do not use these
parameters directly, but transform them into
\begin{eqnarray}  
a_x^2 &=& \frac{a^2(1-\epsilon)^2}{1-\epsilon(2-\epsilon)\cos^2\phi }   \nonumber\\[0.5ex]
a_y^2 &=& \frac{a^2(1-\epsilon)^2}
          {1-\epsilon(2-\epsilon)\cos^2(\phi+\pi/2)}               \\[1ex]
a_{xy} &=& \frac{2-\epsilon(2-\epsilon)
            \left(1+\sin 2\phi \right)}
            {b^2 }- \frac{a_x^2+a_y^2}{a_x^2 a_y^2} \nonumber
\end{eqnarray}
where $\epsilon = 1-b/a$. With these provisions and
assuming for simplicity the centroid to be at
$(0,0)$, the PSF shape in each pixel $(x,y)$ is given by
\begin{equation}
\label{eqn:moff}
F_{\mathrm{PSF}} \:=\: F_{0,\mathrm{PSF}} \,
    \left[\,1+\left(\frac{x^2}{a_x^2}+\frac{y^2}{a_y^2}
    +a_{xy}\,xy\right)(2^{1/\beta}-1)\right]^{-\beta}.
\end{equation}
A similar expression for the PSF was already employed
successfully in crowded field photometry packages such as DAOPHOT
\citep{stet87}, and we simply adopted that concept to
our needs. Its chief benefit lies in the fact that variations in 
position angle over the field, even a complete flip of orientation, 
correspond to secular changes in the $a_x$, $a_y$, $a_{xy}$ parameters.
This fact enables us to use simple bivariate polynomials to describe
the variation of parameters over the field of view, i.e. expressions
of the form
\begin{eqnarray}
a_x (x,y) & = & c_0 + c_1 x + c_2 y + c_3 xy + c_4 x^2 + c_5 y^2 + \dots \nonumber \\
a_y (x,y) & = & d_0 + d_1 x + d_2 y + \dots \\
a_{xy} (x,y) & = & \dots \nonumber
\end{eqnarray}

The actual process to establish a complete PSF model runs as follows:
First the suitable stars are selected. The brightest stars are
modelled individually with a full five-parameter PSF model (Eqn\
\ref{eqn:moff}), with the aim to find a best $\beta$ for the
dataset. Once this is done, $\beta$ is fixed for all subsequent PSF
fits, i.e.\ we do not allow $\beta$ to vary spatially.

In a next step we fit four-parameter models to \emph{all} stars,
using the modified downhill simplex described in detail in the next
section.  This results in a table of PSF parameters at various
positions $(x_i,y_i)$ in the image frame. If the parameters are
consistent with being constant over the frame, or if the scatter is
much larger than any possible trend, the simple average is taken,
otherwise a least-square bivariate polynomial is computed. We have
currently implemented polynomial orders between 1 (bilinear) and 3
(bicubic). The degree which fits best is taken for the final PSF
model, with the additional condition that the \emph{gradient} of the
polynomial should be small in the vicinity of the AGN.  Extremely
ill-fitting stars (and undetected binaries, galaxies etc.)  are
iteratively removed from the table and do not contribute to the
variation fit.

In the example of Fig.~\ref{fig:fish} we plotted position angle and
ellipticity of all usable stars along with a grid of reconstructed
values. The number of stars in the example is high, but not
exceedingly so. The stability of the process allows us to use stars
significantly fainter than the quasar which are of course much more
numerous. In our applications like those presented in section
\ref{sec:obs} we typically find 20 -- 30 or so usable stars per image.

\subsubsection{Lookup table correction}  \label{sec:lut}

For cases where the quality of the PSF determination is critical, i.e.
for data with bad seeing or compact hosts, the analytic representation
of the PSF may be an over-simplification. Without giving away the
advantages of the analytic description, we can apply two second-order
corrections in the form of empiric lookup tables (LUTs):
\begin{equation}
  \label{eqn:lut}
  F_{\mathrm{PSF, corr.}} \:=\: F_{\mathrm{PSF}} + 
  N \left[L_1 (r_n) + L_2 (x,y)\right]
\end{equation}
with $r_n = r / r_{1/2}$ being the normalised radius, $r$ the
elliptical radius as described in Eqn~\ref{eqn:rad} and $N$ a
scaling factor for the LUTs. Here we distinguish between the 
case of azimuthally symmetric errors and that of errors with more 
complicated or no symmetries:

The one-dimensional (radial) LUT $L_1$ contains those corrections that 
show the same symmetry and variation as the model PSF itself. It describes
the intrinsic radial shape difference between the simple analytical
model and the more complicated PSF and can be expressed as an additive
term in Eqn~\ref{eqn:moff1}.

\begin{figure}
  \includegraphics[bb = 47 233 314 484,width=\colwidth,clip]{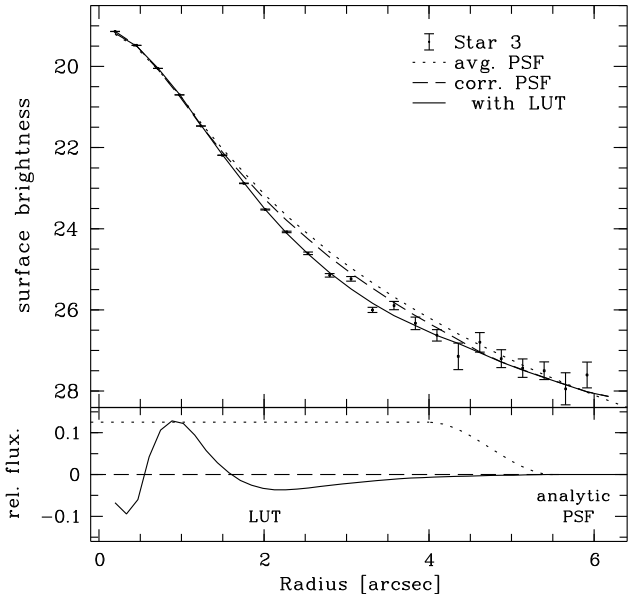}
  \caption{Comparison of profiles with and without radial LUT. As
    example we take star 3 from Fig~\protect{\ref{fig:psf}}. Top panel
    shows profiles of the star (dots), the best-fitting model with LUT
    (full line) and without LUT (dashed).  Bottom panel shows the
    radial lookup table (solid line) in per cent of the total flux,
    together with a scaled transition function $f(r)$ (dotted line)
    defining the outer LUT boundary.}
  \label{fig:lut}
\end{figure}

In practice, $L_1 (r_n)$ is obtained by assessing the residuals of PSF
stars, normalised to unit integral flux, after subtraction of the best
fitting analytic model.  Of those we compute radial profiles spaced in
equidistant fractions of $r_{1/2}$. 

For each radial bin we then average the individual residual profile
value for all stars.  Due to the previous normalisation and azimuthal
averaging, this process is now independent of the spatial PSF
variation. The resulting lookup table $L_1 (r_n)$ can then be used to
correct the symmetric radial errors according to Eqn~\ref{eqn:lut}.
In Fig~\ref{fig:lut} we have done this for the image presented in
Fig~\ref{fig:fish}. The purely analytic model can describe the profile
only up to a certain degree. To improve the fit (most conspicuously needed
between two and four arcseconds) we add the radial LUT, scaled by the
total stellar flux, in a range where it can be determined with high
S/N. Note that the scale of the LUT is linear while the profiles are
plotted logarithmically, the LUT is hence mostly needed in the centre,
not in the wings.

In the next step we apply this global correction to all stellar
models, adapted to their individual model geometry, and again record
the residual images. Averaging these residuals after flux normalisation 
yields a two-dimensional array $L_2 (x,y)$ which is just the desired 
lookup table. To avoid sampling errors, the images should be resampled 
to have the same subpixel centroids.

The quality of both corrections is necessarily a function of the number
of stars available, and of their S/N ratios. In any case, for both the one- and
two-dimensional LUT there exists a radius beyond which Poisson noise
will dominate. The LUTs should be truncated at this radius to avoid
the introduction of additional noise. To avoid artefacts at
the cut-off radius, we apply a smooth transition. For this we define a
transitional annulus [$r_1$:$r_2$] where $\tilde{L}=f(r)L(r)$ with
$f(r)$ a third order polynomial for which holds:
\begin{eqnarray*}
f(r_1) &=& 1\\
f(r_2) &=& 0\\
f'(r_1) &=& f'(r_2) = 0\,.
\end{eqnarray*}

The transition radii are determined interactively as the range where
noise starts to dominate the LUT. An example of the transition
function $f(r)$ is shown in Fig~\ref{fig:lut}. Up to a radius of $4''$
we have $f(r) = 1$, while within the transition annulus $f(r)$
decreases to 0. The effective $\tilde{L}$ is also plotted.

\smallskip

In Fig~\ref{fig:psf} we show the improvement in  PSF fitting with each
successive increase in model complexity. In the top three
rows we plot logarithmic contours of three stars and the best-fitting
models as well as linear contours of the resulting residuals. In the
following rows we successively reduce the model complexity which leads
to an increase in the residual structure as well as in the rms of the
residual as shown in Table~\ref{tab:psfrms}. Taken all corrections
together we now have a high S/N-model of the PSF.

\medskip
In order to estimate the accuracy of the above process, we adopted
the 'leaving one out' method from \citet{duda73}. We repeat the PSF
determination but leave one star out. From the remaining stars we get
a prediction of the PSF parameters at the star's position which is
independent from the star itself. We do this for all the stars and
average the differences between predicted and measured PSF parameters.
If the stars cover the field evenly this will be a good estimate for
the uncertainty of the QSO PSF.

\subsection{Image decomposition}
\label{sec:agn}

\subsubsection{Models}

In order to describe the surface brightness distribution
of QSO host galaxies we have restricted ourselves to the two
most commonly used analytical prescriptions -- an exponential
\citet{free70} law describing early type disc galaxies,
and a \citet{deva48} `$r^{1/4}$'
law describing spheroidals, applicable to elliptical galaxies and 
disc galaxy bulges:
\begin{eqnarray}   
 F_{\mathrm{disc}}(r) &=& F_{\mathrm{disc},0} \exp\left(-1.68\,\frac{r}{r_{1/2}}\right)
 \label{eqn:disc}\\   
 F_{\mathrm{sph}}(r) &=& F_{\mathrm{sph},0} \exp\left[-7.67\,\left(\frac{r}{r_{1/2}}\right)^{1/4}\right]
 \label{eqn:sph}   
\end{eqnarray}   
where the `radius' $r$ is a function of $x$ and $y$:
\begin{equation}
    \label{eqn:rad}
    r^2 \:=\: \frac{1-\epsilon(2-\epsilon)\cos^2(\alpha-\phi)}{(1-\epsilon)2}\left(x^2+y^2\right)\:,
\end{equation}
with $\tan\alpha = y/x$. (Exponential bulges in late-type spirals are
currently not modelled as these are not known to harbour significant
nuclear activity).  Thus, each galaxy model contains four independent
parameters: The semimajor axis $a$ for which holds that $r(a) =
r_{1/2}$; the ellipticity $\epsilon = 1-b/a$; the position angle of
the major axis, $\phi$; and the total flux $F = \int F(r) \: \mathrm{d} r\:\mathrm{d} \alpha$.
Notice that we avoid to use the ill-constrained central surface
brightness as a fit parameter. It is well known that the determination
of effective radius and central surface brightness is strongly
degenerate in the presence of measurement errors
\citep[e.g.][]{abra92}, and that the total flux $F_0$ is much better
constrained than either of these parameters. This issue will be
addressed again in Sect.~\ref{sec:sim} below, in particular in
Figs~\ref{fig:test1} and \ref{fig:test2}.

To summarise, a typical model will contain either five or nine
parameters: four for each galaxy component, plus a point source
scaling factor for the AGN. However, we have also implemented an option
to keep individual parameters at a fixed value, 
so that the above numbers give the \emph{maximum} number of parameters.

\subsubsection{Convolution}

\begin{figure}
  \includegraphics[width=\colwidth,clip]{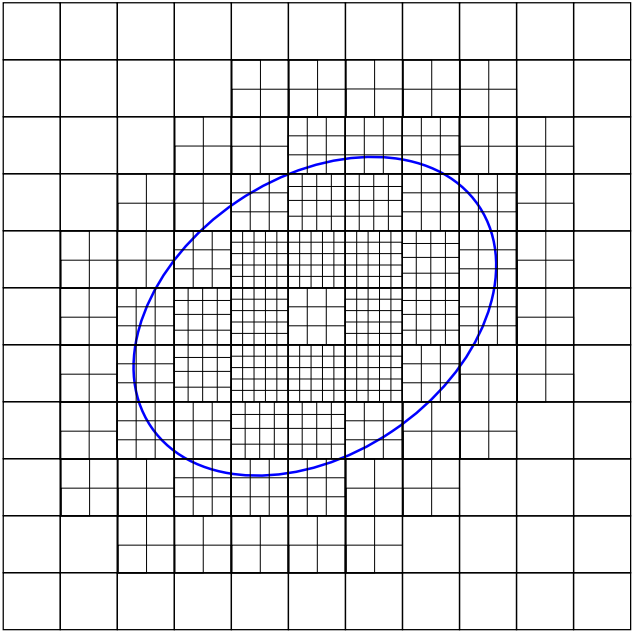}
  \caption{ Illustration of the adaptive
    subxpixelling. Each pixel with a gradient larger then a threshold
    value is divided into subpixels. These subpixels itself are
    divided as long the gradient is still too large. The size of the
    original pixels is maintained in the outer parts. The ellipse
    represents this object's half-light isophote.}
  \label{fig:grid}
\end{figure}
 
Although both PSF and galaxy are represented by analytic functions,
the nonzero ellipticities demand that the convolution be evaluated
numerically. In numerical convolution, sampling plays an important
role: strictly speaking, we have to distinguish between (a) the
function value at given $x,y$; (b) the PSF-convolved value; (c) the
image value sampled into a rectangular pixel grid.  These values will
be similar only in areas of small gradients in the surface brightness
distribution; close to the centre, the galaxy light has to be sampled on
a much finer grid in order to avoid large numerical errors.
On the other hand, a highly oversampled pixel grid leads to
a substantial increase in computing time and is therefore
inefficient. It is also not required in the outer regions.

Our adopted solution uses the local gradient in the unconvolved
image to adjust the degree of oversampling, as illustrated in
Fig 4. This adaptive subpixel grid is determined at the beginning
of each fitting subprocess (see below). Whenever the model parameters
change substantially, the grid is recomputed and the fitting process
is resumed with this new grid.

\subsubsection{The fitting process}

The model parameters are iteratively adjusted by minimizing $\chi^2$
with the downhill simplex method \citep{pres95}.  Here, the $\chi^2$
values are based on variance frames associated with each image, which
may also contain information about regions that are to be left out of
the fitting process.

In order to accelerate and stabilise the minimization, the parameter
space is transformed to achieve `rounder' $\chi^2$ valleys. We use
the following transformation recipes: $\epsilon$ is substituted by
$\epsilon' \equiv \log (1-\sqrt{1-\epsilon^2})$, and $F$ is
substituted by $F' \equiv \log (F + O)$.  As a byproduct, this
transformation automatically ensures that F has a lower acceptable
bound $-O$. Note that $O=0$ (i.e.\ demanding $F > 0$) is not always a
good choice; in the case of a faint or undetectable host galaxy and in
the presence of noise, slightly negative values of $F$ must be
permitted.

A crucial part of the algorithm is its subdivision into successive
minimization substeps in order to avoid trapping in local minima.
Whenever a $\chi^2$ minimum is found, the process is restarted at the
same location in parameter space, probing the environment for a
further decrease in $\chi^2$-values. Additional restarts are launched
when the change in parameters requires a reevaluation of the
subpixelling grid. Only when even repeated restarts yield no
improvement in $\chi^2$, the entire process is considered to
have found a global minimum. This way we can usually avoid to be
trapped in shallow local minima or regions of small curvature. 

Fitting the full set of nine parameters is only useful for data with
excellent spatial resolution, providing significant independent
constraints for AGN, disc, and bulge components. There are various
ways to reduce the number of fitting parameters; besides fitting just
one galaxy model, we have included an option to keep parameters at a
fixed value. This is useful e.g.\ in the analysis of multicolour data
where certain structural parameters might be well-constrained in one
dataset (e.g., HST) which then can be used to increase the modelling
fidelity of images taken in other bands.

\begin{table*}
  \caption{\label{tab:sim}Overview over our simulations. The input parameters
    are total counts $F$ (in units of detector photoelectrons) and
    half-light radii $r$. Corresponding absolute
    magnitudes $M$ and linear radii $r$ [kpc] are also listed
    for comparison. For details see text and Table~\protect{\ref{tab:exptime}}.}

\begin{tabular}{lrrrrrrrrrr}
\hline\noalign{\smallskip}
Dataset &$F_\mathrm{nuc}$      & $F_\mathrm{disc}$      & $F_\mathrm{sph}$      &
         \multicolumn{1}{r}{$r_{\mathrm{disc}}$}   & \multicolumn{1}{r}{$r_{\mathrm{sph}}$ } &
         $-M_{R,\mathrm{nuc}}$ & $-M_{R,\mathrm{disc}}$ & $-M_{R,\mathrm{sph}}$ &
         $r_\mathrm{disc}$     & $r_\mathrm{sph}$       \\
        &$10^5$ [e$^-$]        & $10^5$ [e$^-$]         & $10^5$ [e$^-$]        &
         [arcsec]              & [arcsec]               &
                               &                        &                       &
         [kpc]                 & [kpc]                  \\
\noalign{\smallskip}\hline\noalign{\smallskip}
 med $z|_s$ &10       & --      & 2.5     & --       & 1.3      & 24.5 & -- & 23.0       & -- & 10.0 \\
 med $z|_m$ &11--7.8  & --      & 11--0.8 & --       & 1.2--4.1 & 24.5 & -- & 22.0--24.5 & -- & 8.7 \\
 med $z|_e$ &10       & --      & 2.5     & --       & 1.3--10.8& 24.5 & -- & 23.0       & -- & 10.0--78.4 \\
 low $z|_M$ &10       & 1.0--20 & 1.0--20 & 6.0      & 3.0      & 24.2 & 21.7--25.0 & 20--22.5 & 5.0 & 2.5 \\
 low $z|_R$ &10       & 1.0     & 1.0     & 3.0--8.0 & 1.5--6.0 & 24.2 & 21.7 & 21.7 & 2.5--6.8 & 1.3--5.0\\
\noalign{\smallskip}\hline
 \end{tabular}
\end{table*}

\section{Simulations}
\label{sec:sim}

To test the reliability of the AGN decomposition process, we
constructed extensive sets of simulated galaxies. As the multitude of
instruments and objects prevents a test for the full range of possible
data, we limit the test to two rather different sets which both
closely resemble certain observational data recently obtained by us.
One the one hand, we consider a set of low redshift AGN observed with
a 1.5~m telescope.  On the other hand, we consider the case of medium
to moderately high redshift QSOs (up to $z\simeq 1$), observed with a
4~m class telescope.  These simulations resemble the examples
described in Sect.\ \ref{sec:sam} of this paper. These two simulated
datasets will henceforth be referred to as `low $z$' and `med $z$'.
Input properties are listed in Table~\ref{tab:sim}.

We have thus constructed a test bed for two very different
configurations.  The low-redshift objects were created using various
combinations of three components (disc, spheroid and a nuclear
source), and among these objects we expect to find and retrieve all
Hubble types.  For the medium and high redshift data we expect
elliptical galaxies to dominate the host galaxy population. In this
case the objects are compounds of only a spheroidal and a nuclear
component, and we attempt no more than reclaiming the properties of
these two components, concentrating on  luminosities and scale
lengths. In this paper we do not investigate the influence of
inclination on the decomposition process.

Both simulated sets were created using the same radial profiles and
isophotal shapes that we used to compute the model galaxies during the
fitting process. To account for observational errors we added
artificial shot noise. The sets were then treated in the same way as
real observational data. In order to avoid confusion between errors in
the modelling of the spatial PSF variations and the fitting of galaxy
and AGN, we assumed the PSF to be shift-invariant.

\subsection{Medium-redshift simulations}
\label{sec:medsim}

We start with the medium-redshift simulations as these were fitted
with the conceptually simpler two-component models.  The first subset
contains images of only a single galaxy, but `observed' numerous
times, i.e.\ with several different noise realisations, and with
different centroid positions with respect to the pixel grid (dataset
`med $z|_s$', for `single redshift', in Table~\ref{tab:sim}).  The
input galaxy is a typical bright elliptical galaxy with half-light
radius $r_{1/2}=10$~kpc, an absolute luminosity of $M_R =-{24.5}$ at a
redshift of $z=0.6$, with a nucleus four times brighter than the host
galaxy.

\begin{figure}
\setlength{\unitlength}{1cm}
\begin{picture}(6,9)
\put(0.6,8.5){Composite}
\put(3.4,8.5){Nucleus}
\put(6.2,8.5){Galaxy}
\put(0.6,4.0){Composite}
\put(3.4,4.0){Nucleus}
\put(6.2,5.5){Disk}
\put(6.2,2.5){Spheroid}
\put(0.2,6.1){\rotatebox{90}{z = 0.6}}
\put(0.3,7.2){\line(0,1){1.0}}
\put(0.2,7.2){\line(1,0){0.2}}
\put(0.2,8.2){\line(1,0){0.2}}
\put(0,7.6){\rotatebox{90}{5$''$}}
\put(0.5,6.0){%
\includegraphics[bb = 100 100 160 160,height=\dantab,clip]{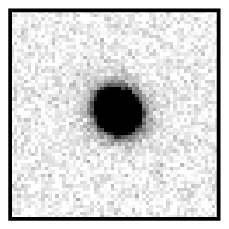}}
\put(3.3,6.0){%
\includegraphics[bb = 100 100 160 160,height=\dantab,clip]{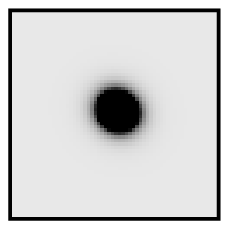}}
\put(6.1,6.0){%
\includegraphics[bb = 100 100 160 160,height=\dantab,clip]{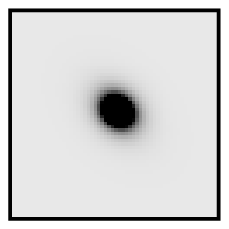}}
\put(0.2,1.6){\rotatebox{90}{z = 0.02}}
\put(0.3,2.8){\line(0,1){0.6}}
\put(0.2,2.8){\line(1,0){0.2}}
\put(0.2,3.4){\line(1,0){0.2}}
\put(0,2.9){\rotatebox{90}{10$''$}}
\put(0.5,1.5){%
\includegraphics[bb = 100 100 201 201,height=\dantab,clip]{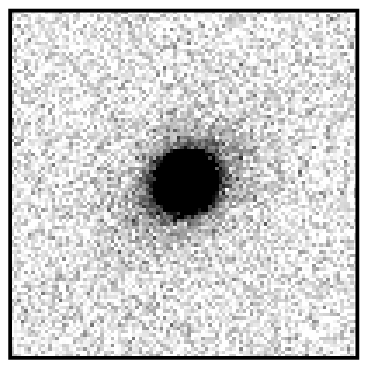}}
\put(3.3,1.5){%
\includegraphics[bb = 100 100 201 201,height=\dantab,clip]{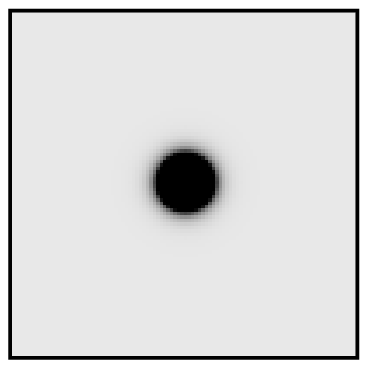}}
\put(6.1,3){%
\includegraphics[bb = 100 100 201 201,height=\dantab,clip]{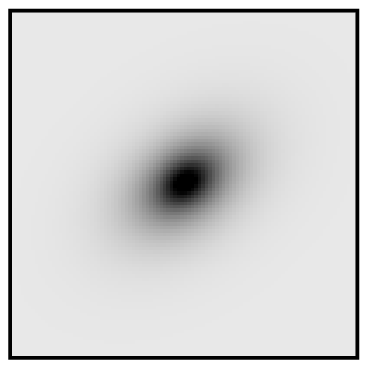}}
\put(6.1,00){%
\includegraphics[bb = 100 100 201 201,height=\dantab,clip]{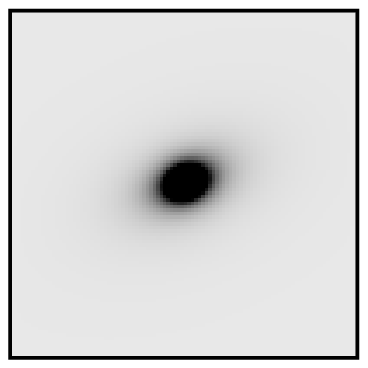}}
\end{picture}
  \caption{Two example model galaxies. In the top row we show
    the composite model and the single components of the `med $z|_s$'
    model,in the bottom row the same is done for the `low $z|_M$'
    model with the weakest host. Scales and cuts are held constant for each redshift.}
  \label{fig:models}
\end{figure}

To compute realistic flux and background levels, we used the exposure
time calculator for the ESO-NTT and its multi-mode instrument EMMI,
assuming a pixel size of $0\farcs27$ and a total exposure time of
500~s per simulated image.  In order to specify the background level,
we assumed the data to be obtained in the $V$ band.  The adopted PSF
has a width of $0\farcs8$ FWHM, compared to $r_{1/2}=1\farcs33$ for
the galaxy.

\begin{figure}
  \includegraphics[bb = 71 573 337 772,width=\colwidth,clip]{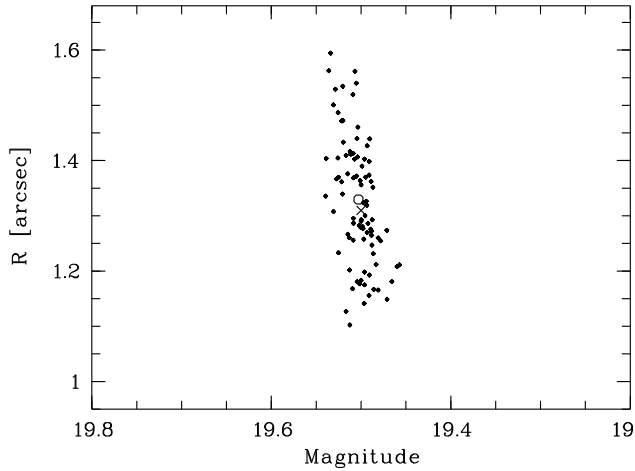}
   \caption{Results for the `med $z|_s$' simulation, featuring 
   different noise realisations and subpixel locations. 
   Each dot represents the result of fitting one particular image. 
   The circle indicates the average of the fitted values, and
   the cross denotes the `true' input value. 
   The nucleus is brighter than the host galaxy by 1.5~mag. 
   The scatter of extracted parameter
   values (1$\sigma$) is 0.02 for the magnitude and 9 per cent for the
   radius.}  
   \label{fig:test1}
\end{figure}

Fitting the simulated images of this dataset, we found that we are
able to reclaim the original host galaxy magnitude with an uncertainty
of only 0.02 (1$\sigma$). This is shown in more detail in
Fig~\ref{fig:test1}, which also illustrates the well-known fact that
the half-light radius is less accurately recovered.  However, with an
uncertainty of 9 per cent in $r_{1/2}$ we are still able to give a solid
estimate of the galaxy size, even at this redshift and with a host
galaxy only slightly more extended than the PSF.

In the second dataset (`med $z|_m$', for `multiple redshifts'), we
placed the same galaxy at four different redshifts ($z=0.1$, 0.2, 0.4,
1.2) and changed the galaxy flux such that the ratio nucleus/galaxy
takes three different values (10:1, 4:1, 1:1).  To enable a fair
comparison, the exposure time in each case was adjusted to yield the
same S/N for all redshifts (cf.\ Table \ref{tab:exptime}), and the
underlying spectrum was assumed to be flat, i.e.  we have the same
luminosity in all the spectral bands. This latter assumption is
obviously unphysical, but acceptable for our illustration purposes as
the main free input parameters are the nuclear flux and the
nuclear/host flux ratio.

\begin{figure}
  \includegraphics[bb = 71 573 355 772,width=\colwidth,clip]{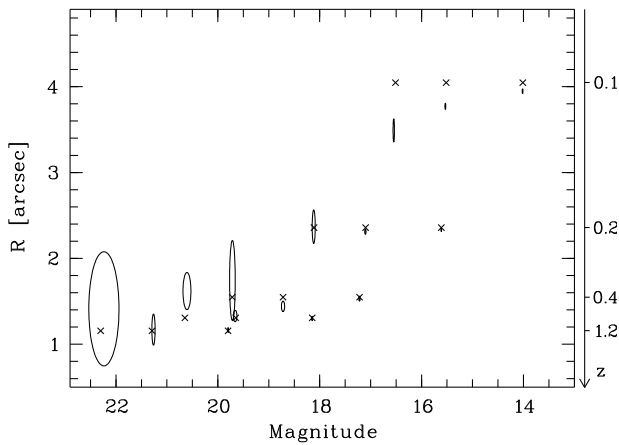}
  \caption{Results for the `med $z|_m$' simulations, involving
  four different redshifts. Crosses represent the
  input values, and the ellipses approximately delineate the scatter 
  of the extracted parameter values, with a minor semiaxis of 2$\sigma$ 
  in magnitude and a major semiaxis of 1$\sigma$ in radius. 
  The magnitude of the nuclear component is equal to that of the brightest 
  host galaxy at each redshift. Values are given
  in arcseconds and $R$-band apparent magnitudes.}  
  \label{fig:test2}
\end{figure}

For each configuration we generated images with several different
noise realisations and fitted those independently. The results show
clearly and not surprisingly that the accuracy of recovering the input
parameters depends on redshift (see Fig~\ref{fig:test2}).  But even in
the case of the most unfavourable redshift, $z=1.2$ and the highest
nuclear/host ratio, the reconstructed host galaxy luminosity has an
rms scatter of less than 0.15~mag (1$\sigma$). Again, the half-light
radii are less accurately determined. Ellipticity and position angle
were held constant during these simulations ($\epsilon = 0.4$, PA =
20$^\circ$). Fitted values agreed on average with the model
values with scatters below 2 per cent in $\epsilon$ and $3^\circ$ in
PA except for the faintest models where it rose to 20 per cent and
$8^\circ$ for the faintest. This held for all redshifts except the
lowest where the scatter was significantly smaller.

\subsection{Influence of external parameters}

\begin{figure}
  \includegraphics[width=\colwidth,clip]{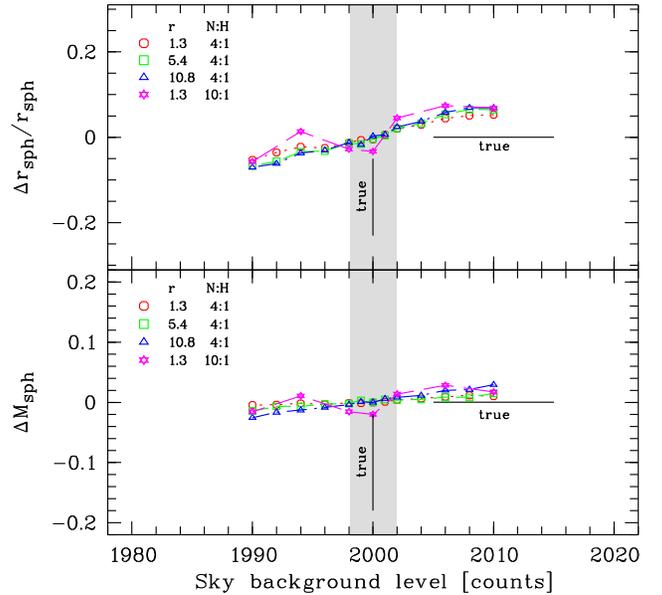}
   \caption{Dependence of recovered radii (top) and magnitudes of the
     galaxy (bottom) on the on the accuracy of the background
     determination. Instead of using the true value (marked with a
     vertical bar) we used offset values for the sky background to fit
     the data. The shaded area is the range of the typical background
     uncertainty.}
   \label{fig:backtest}
\end{figure}

\begin{figure}
\includegraphics[width=\colwidth,clip]{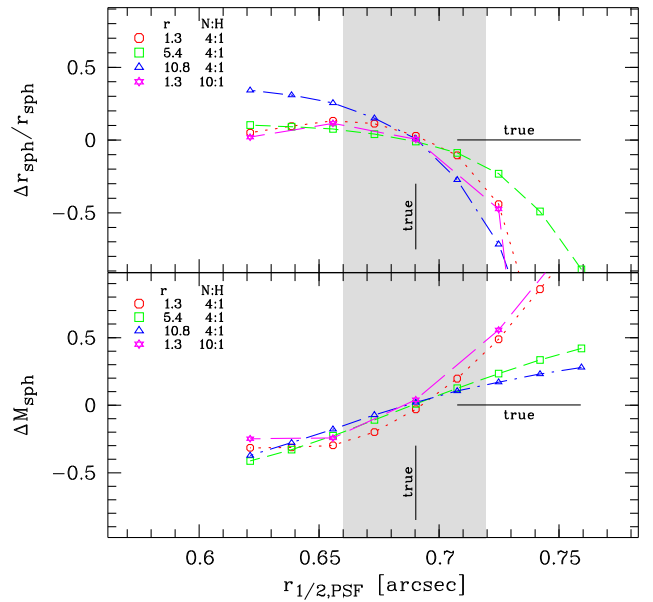}
   \caption{Dependence of recovered radii (top) and magnitudes of the
     galaxy (bottom) on accuracy of the half-light radius of the PSF.}  
   \label{fig:rtest}
\end{figure}

In the simulations we assumed that we know the true value of the sky
background and the PSF parameters. In reality these are afflicted with
uncertainties. To test their influence we created a set of models
(`med $z|_e$', for `external') similar to the `med $z|_s$' simulations
(which have $z=0.6$ and a nucleus four times brighter than the host)
but with three different galaxy radii ($r_{1/2}=1\farcs33, 5\farcs4,
10\farcs8$) in order to resemble observations typical for our group.

These models were fitted using deliberately wrong values (one at a
time) for the sky background, which is notorious for influencing the
results, and the PSF half-light radius $r_{1/2}$ which appeared to be
the most critical parameter. For each configuration we generated a
number of different noise realizations and computed the average values
of recovered host galaxy radius and magnitude.

In Figs~\ref{fig:backtest} and \ref{fig:rtest} the results can be
compared. While in the typical range of errors the uncertainties
induced by an uncertain background are almost negligible for the
magnitude and below 5 per cent for the radius, the accuracy of the
determined PSF half-light radius is essential. Errors are as large as
0.5 magnitudes or 50 per cent for the radius here.

\begin{figure}
  \includegraphics[bb = 71 573 337 772,width=\colwidth,clip]{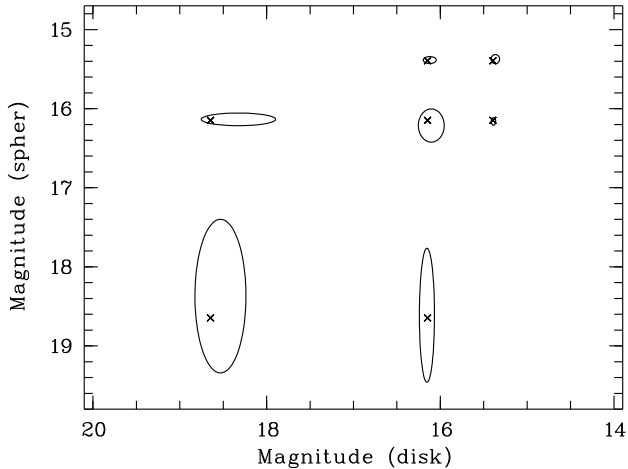}
  \caption{Results for the `low $z|_M$' simulation, showing the accuracy
    of reclaiming component luminosities in three-component fits. 
    The error ellipses have semiaxes of 2$\sigma$ in magnitude.  
    The nuclear component has a magnitude of 16.15 for
    all objects.}
  \label{fig:test3}
\end{figure}

We also created models which have nuclei 10 times brighter than the
host galaxy, similar to the `worst case' in the $z=1.2$ `med $z|_m$'
models. Fitting those with offset values for the background and the
PSF half-light radius, we find the systematic errors to be comparable
to those of the `typical' models (see Figs~\ref{fig:backtest} and
\ref{fig:rtest}). In the background test the larger statistical errors
introduced by the fitting process (see Sect.\ \ref{sec:medsim} and
Fig~\ref{fig:test2}) become visible.  This is less visible in the PSF
half-light radius test, as the different nuclear-to-host relation play
a less prominent role. Here, the error of the nucleus dominates as long as
it is notably brighter than the galaxy.

\subsection{Low-redshift simulations}

For well resolved AGN host galaxies a three-component fit may be more 
appropriate. In our low-redshift sample, the host galaxies are of
all Hubble types and their morphology can be easily resolved even with
small telescopes as we will show in the next section. To test the
three component fitting, we generated a dataset to match those
observations.

We simulated galaxies with both a disc and a spheroid and a
bulge-to-total ($b/t$) flux ratio between 0.1 and 0.9. The ratio
between nuclear and galactic light was varied between 5:1 and 1:4 (set
`low $z|_M$' for `magnitude variation' in Table~\ref{tab:sim}). The
half-light radii were set to typical values found in our
observed sample. All galaxies are azimuthally
symmetric, no late-type features like bars or spiral arms were added.

Note that the simulations were designed to match the observations in
integrated flux and apparent radii. Values in the table are given for
a template observation of 840~s exposure time (on a 1.5~m telescope)
and a redshift of 0.019, which was also used to compute the level of
noise of 800 e$^-$/Pixel at a pixel size of 0\farcs39. We assumed a
seeing of 1\farcs6 (FWHM), which is rather poor but unfortunately was
typical for our observations.

Fig~\ref{fig:test3} shows the results of the fits. The
property dominating the uncertainty is the flux ratio between nuclear
component and the galaxy (moving from lower left
to upper right in Fig~\ref{fig:test3} decreases this ratio). 
The bulge magnitude is more affected by this than the disc magnitude, 
which is easily explained by the lower half-light radii of the 
bulge component, which is thus harder to be distinguished from 
the nucleus. The 1$\sigma$ uncertainty grows from 0.03~mag for a 
ratio of 1:2 (nuclear to spheroidal flux) to 0.46~mag for a ratio
of 10:1); the corresponding values for the disc component are
0.02~mag at 1:2 and 0.2~mag at 10:1.

\begin{figure}
  \includegraphics[bb = 71 573 337 772,width=\colwidth,clip]{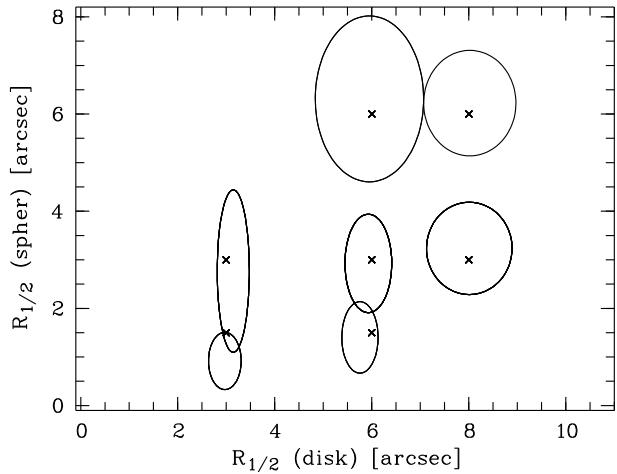}
  \caption{Results for the `low $z|_R$' simulation, featuring three-component
    models with different half-light radii.
    Error ellipses have semiaxes of 2$\sigma$
    of the radius (in arcseconds). All objects have a nuclear to
    total galactic flux ratio of 5:1 and a $b/t$ of 0.5. 
    Radii are given in arcseconds.}
  \label{fig:test4}
\end{figure}

In order to probe how well galaxy sizes
can be recovered with these multicomponent fits, we varied the 
radii of both components between 1\farcs 5 and 6\farcs 0 (bulge) 
and 3\farcs 0 and 8\farcs 0 (disc) but left the fluxes unchanged,
with the flux ratio set to the worst-case value of 10:1 for each
component (dataset `low $z|_R$', for `radius variation'). 
Fig~\ref{fig:test4} shows that even when the nucleus dominates 
over the galaxy, the relative error in the determination of the
half-light radius is reasonably low ($\sim 5$ per cent for the disc 
and $\sim 20$ per cent for the spheroidal component). 

No special simulations were done for ellipticity and position angle.
Within the above simulations, where both had constant values
($\epsilon = 0.33$, $\phi_\mathrm{sph}=22^\circ$,
$\phi_\mathrm{dis}=37^\circ$), they were on average fitted well with
scatters below 2 per cent in $\epsilon$ and $2^\circ$ in PA.  Again for
the faintest galaxies the scatter rose to 6 per cent and $4^\circ$ and
as high as 25 per cent and $6^\circ$ if the galaxy component was
hidden by both a bright nucleus and a bright second galaxy component.
We did not do specific simulations for other values of $\epsilon$ and
$\phi$, but tests suggest that for larger values of $\epsilon$ both are
determined even better, while for smaller values no large differences
are expected, as the above case is already almost circular.

We conclude by stating that our simulations have yielded 
encouraging results. Total host galaxy luminosities can be reclaimed
with high fidelity, and although half-light radii are less accurately
constrained, there is no evidence for systematic errors.
Recall that noise level, pixel sampling, and in particular seeing 
in these simulations were matched to our already existing data.
It would be easy to design additional datasets obtained under 
better conditions, in which case a substantial improvement of
measurement accuracy can be expected. We stress, however, the 
importance of individually tailored simulations in order to 
assess the potential and limitations of each observed dataset.

\begin{table}
  \caption{ Redshifts, apparent nuclear magnitudes, exposure times 
    and resulting sky background contribution adopted as input for the  
    simulations.  \label{tab:exptime}}
  \begin{tabular}{l|lrrrr}
     \hline\noalign{\smallskip}
    Set      &$\:\:\: z$   & $V$  &\multicolumn{1}{c}{$t_{\mathrm{exp}}$}    &\multicolumn{1}{c}{BG}       \\
             &             &      & \multicolumn{1}{c}{[s]      } &   [e$^-$/Pixel]\\
     \noalign{\smallskip}\hline\noalign{\smallskip}
     low $z$ & 0.02 & 14.0 &  840  & 800  \\
     med $z$ & 0.1  & 14.3 &  10   & 40   \\
     med $z$ & 0.2  & 15.9 &  44   & 176  \\
     med $z$ & 0.4  & 17.5 &  200  & 800  \\
     med $z$ & 0.6  & 18.4 &  500  & 2000 \\
     med $z$ & 1.2  & 20.1 &  3100 & 12400\\
     \noalign{\smallskip}\hline
  \end{tabular}
\end{table}

\section{Example applications} \label{sec:obs}

\subsection{Samples and observations}  \label{sec:sam}

As a first test case with real data, we have investigated two
statistically complete samples of quasars and Seyfert~1 galaxies. The
objects form subsamples of the Hamburg/ESO survey
\citep[HES,][]{wiso00} and constitute quasars and Seyfert~1 galaxies
with redshifts $z<0.35$ resp. $0.01<z<0.05$.  Typical nuclear absolute
magnitudes are around $M_H \simeq -27$ (medium redshift) and $M_B
\simeq -21$ (low redshift).

All 13 low redshift objects were observed in the $R$ band using the
ESO/Danish 1.54~m telescope on La Silla and its multi-purpose
instrument DFOSC.  The seeing during the three nights of observation
was rather poor (1\farcs3--1\farcs8), but due to their low redshifts
all of our objects were spatially well resolved.

The 42 med-z quasars were observed in $H$ band during two nights using
the ESO NTT telescope with the SOFI camera with a seeing of
0\farcs4--0\farcs8. 

The images were reduced with standard procedures (debiasing, flatfielding) 
and flux-calibrated using standard star sequences taken in the same nights.
For the infrared images the sky background / bias was determined using
stacks of dithered on-target images. For analysis we shifted the images
to same object centroids using subpixeling and coadded the images.

\subsection{Modelling}  \label{sec:fit}

The fitting of the data followed the procedure laid out in sections
\ref{sec:psf} and \ref{sec:agn}. The PSF determination for the low
redshift observations was straightforward, as with the large field of
view of the DFOSC detector (13\fmins 3 $\times$ 13\fmins 3) always a
large number of stars were available in the image, at least 20 or 30.
The field of view of SOFI is notably smaller (4\fmins 6 $\times$
4\fmins 6) but as the images are also deeper the number of usable
stars is again typically 20 or 30 and always greater than 12.
Depending on the image, a second or third order polynomial could
usually represent the variation to sufficient accuracy.
Figs~\ref{fig:fish} and \ref{fig:psf} were actually created from this
data.

\begin{figure}
\setlength{\unitlength}{1cm}
\begin{picture}(6,6)
\put(0.7,5.6){Data}
\put(3.5,5.6){Profiles}
\put(6.2,5.6){Residual}
\put(0.7,2.6){Data}
\put(3.5,2.6){Profiles}
\put(6.2,2.6){Residual}
\put(0.0,3.3){\rotatebox{90}{HE\,0444--3449}}
\put(0.3,3.0){%
\includegraphics[bb = 91 642 244 771,height=\proftab,clip]{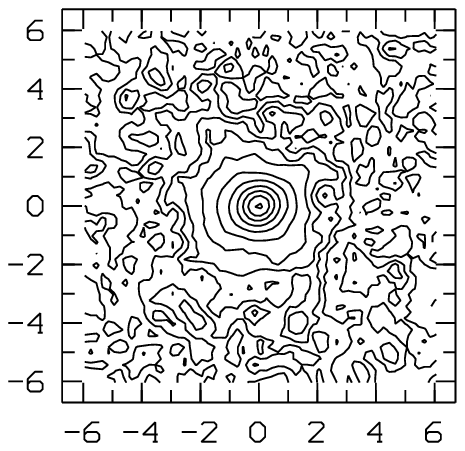}}
\put(3.1,3.0){%
\includegraphics[bb = 91 642 244 771,height=\proftab,clip]{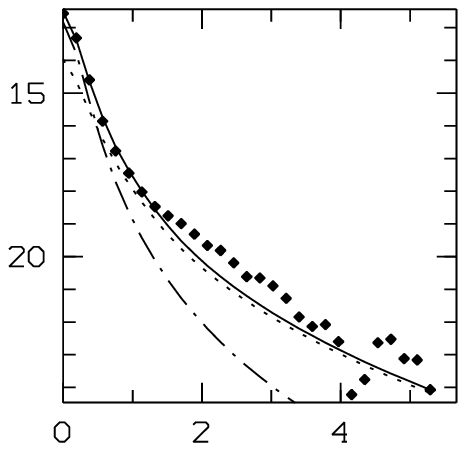}}
\put(5.9,3.0){%
\includegraphics[bb = 93 93 162 162,height=\proftab,clip]{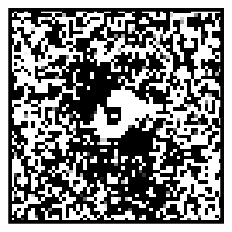}}
\put(0.0,00.3){\rotatebox{90}{HE\,0914--0031}}
\put(0.3,00){%
\includegraphics[bb = 91 642 244 771,height=\proftab,clip]{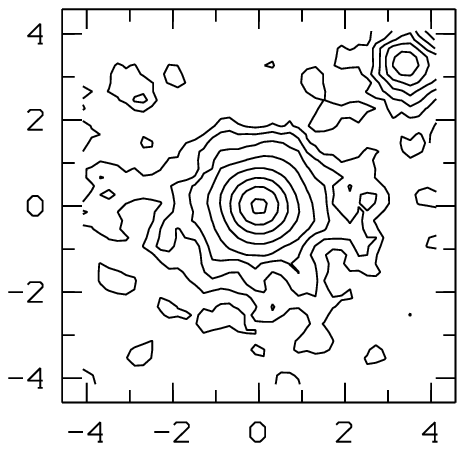}}
\put(3.1,00){%
\includegraphics[bb = 91 642 244 771,height=\proftab,clip]{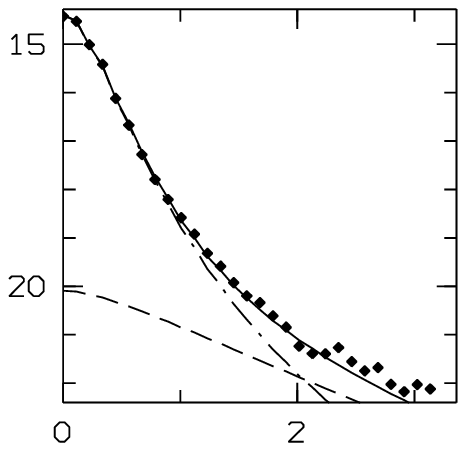}}
\put(5.9,00){%
\includegraphics[bb = 95 95 143 143,height=\proftab,clip]{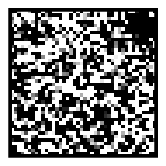}}
\end{picture}
  \caption{Two examples from the medium redshift sample at $z=0.18$
    (top) and $z=0.32$.  Left column: contour plots, with contour
    spacings of one magnitude. Middle: Azimuthally averaged profiles,
    with dots representing the observed data and the solid line
    denoting the overall fit.  Dashed, dotted, and dashed-dotted lines
    correspond to the disc, spheroidal, and nuclear model components,
    respectively.  Right: residual images after subtraction of
    all model components.  Cut levels are set at $\pm$ the value of
    the outermost isophotes shown in the contour plots. White/black
    areas indicate regions where the model is brighter/fainter,
    respectively, than the data.  Units are arcseconds in the contour plots and mag$/\sq''$
    against arcseconds in the profiles.}
  \label{fig:datamed}
\end{figure}

\begin{figure}
\setlength{\unitlength}{1cm}
\begin{picture}(6,6)
\put(0.7,5.6){Data}
\put(3.5,5.6){Profiles}
\put(6.2,5.6){Residual}
\put(0.7,2.6){Data}
\put(3.5,2.6){Profiles}
\put(6.2,2.6){Residual}
\put(0.0,3.3){\rotatebox{90}{HE\,1017--0305}}
\put(0.3,3.0){%
\includegraphics[bb = 91 642 244 771,height=\proftab,clip]{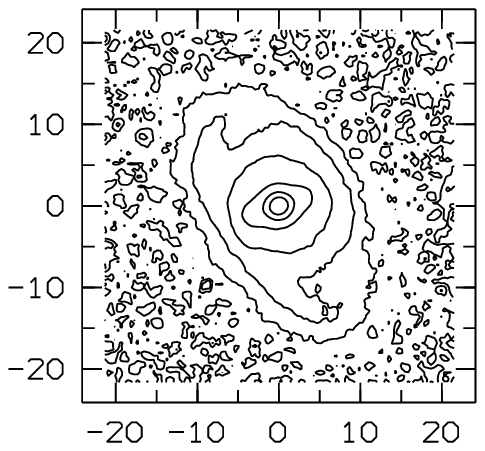}}
\put(3.1,3.0){%
\includegraphics[bb = 91 642 244 771,height=\proftab,clip]{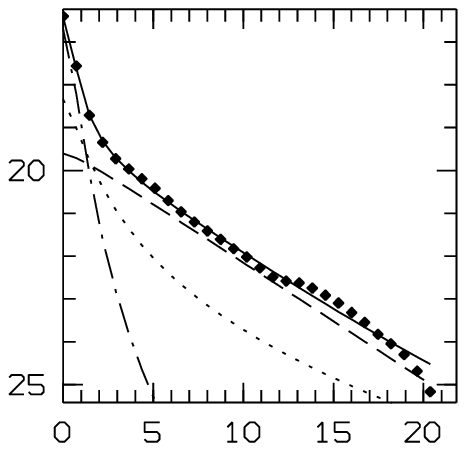}}
\put(5.9,3.0){%
\includegraphics[bb = 87 87 209 209,height=\proftab,clip]{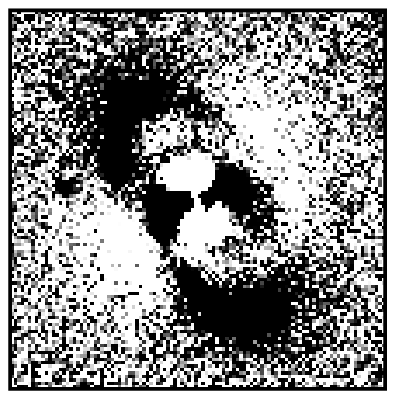}}
\put(0.0,00.3){\rotatebox{90}{HE\,1348--1758}}
\put(0.3,00){%
\includegraphics[bb = 91 642 244 771,height=\proftab,clip]{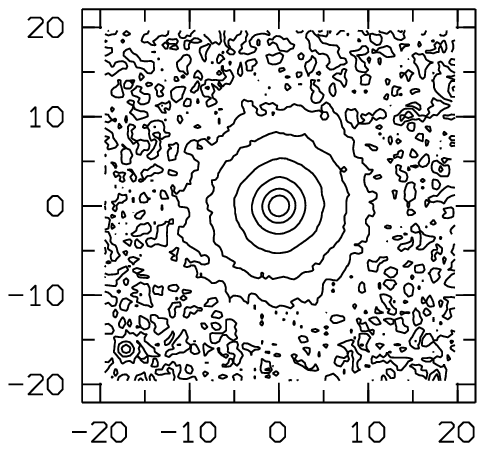}}
\put(3.1,00){%
\includegraphics[bb = 91 642 244 771,height=\proftab,clip]{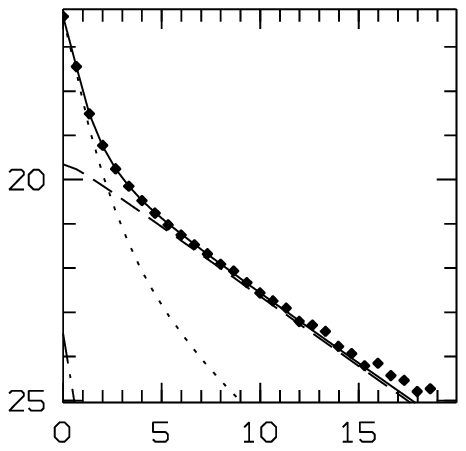}}
\put(5.9,00){%
\includegraphics[bb = 88 88 200 200,height=\proftab,clip]{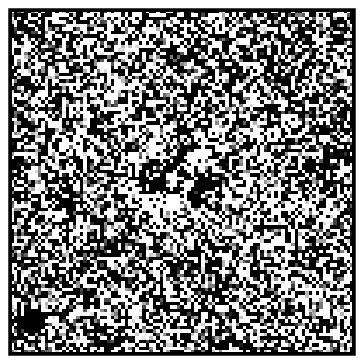}}
\end{picture}
  \caption{Two examples from the low redshift sample.
    Notation is as in Fig~\protect{\ref{fig:datamed}}}
  \label{fig:datalow}
\end{figure}

Some preparatory work before the host galaxy modelling involved
fine-tuning of the local sky background near each AGN using growth
curves, and masking all features in the frames that clearly do not
belong to the object. The maximum fitting radius was set to an ellipse
containing 99.5 per cent of the total object flux.  The contour plots
in Figs~\ref{fig:datamed} and \ref{fig:datalow} have been made just
large enough for this ellipse to fit in.

Good initial parameter estimation is very important to avoid local
$\chi^2$ minima located at parameter combinations very different from
those near the global minimum.  At least with the simplex method it is
difficult to leave such a minimum, once trapped in it. We estimated
initial parameters in the following way: We first determined the
isophotal shape of the disc component (nearly always the most extended
component) by fitting ellipses to the outermost isophotes.  The scale
length and total flux was then obtained from fitting an exponential
law to the outermost part of the surface brightness profile.  The
determination of the bulge parameters was done likewise, but using the
original image with a convolved disc component subtracted.  Finally
the remaining central flux was attributed to the nucleus.  If any of
these steps led to unsatisfactorily strong residuals, the process was
repeated in a different order (first spheroid, then disc).  The
parameter values obtained from this procedure were used as initial
guesses, enabling us to start the full three component fitting for the
low redshift sample as well as to decide which galactic component to
use in a two component fit for the medium redshift sample. In only
five cases the resolution of the medium redshift images was good
enough to allow a three component fit.

The quality of each fit was investigated manually by checking the
resulting profiles, residual images (such as those shown in
Fig~\ref{fig:datamed}), the sequence of $\chi^2$ values, and the
plausibility of the parameters obtained.  If a fit was not
satisfactory, i.e.  leading to strong residuals or to discrepancies
with the object's profile which could not be attributed to prominent
features in the galaxy, we spent more effort in estimating good
initial conditions, or imposed additional constraints in the form of
boundary conditions to ensure that the fitting results corresponded to
physically meaningful components.  For a few objects the three
component fit suggested that two components might be sufficient to
model the light distribution.  For these we repeated the fit with only
two components and selected that if the fit was satisfactory. We
comment on a few such cases below.

For the three low redshift objects where just a nuclear and a disc
component were required we estimated an upper limit for the bulge
luminosity by adding compact artificial spheroids ($r_{1/2}= 1, 5,
10$~kpc) with successively decreasing fluxes. These images were fitted
with both a nucleus plus disc and with a three components model.  The
faintest spheroid for which a three component fit is preferred (i.e.
has a significantly lower reduced $\chi^2$) is then taken as
limit for the detection of a spheroid in that object.  We did not
determine limits for the bulge size, as the sensitivity on the size
drastically reduces towards low galaxy fluxes (see above).

For one object, HE\,1348$-$1758, which did not show any nuclear
component, an upper limit was estimated in a similar fashion by adding
an artificial nucleus.

\begin{figure}
  \setlength{\unitlength}{1cm}
\begin{picture}(6,6)
\put(0.7,5.6){Data}
\put(3.5,5.6){Profiles}
\put(6.2,5.6){Residual}
\put(0.7,2.6){Data}
\put(3.5,2.6){Profiles}
\put(6.2,2.6){Residual}
\put(0.0,3.3){\rotatebox{90}{HE\,1017--0305}}
\put(0.3,3.0){%
\includegraphics[bb = 91 642 244 771,height=\proftab,clip]{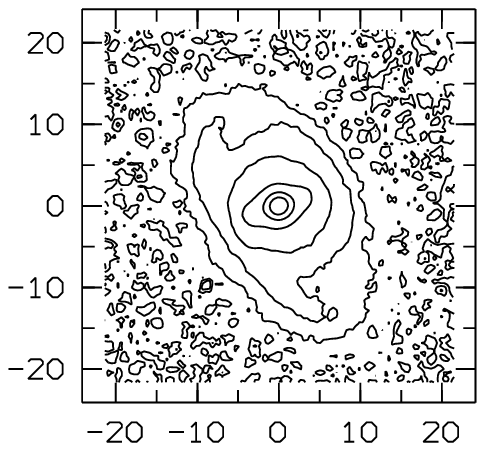}}
\put(3.1,3.0){%
\includegraphics[bb = 91 642 244 771,height=\proftab,clip]{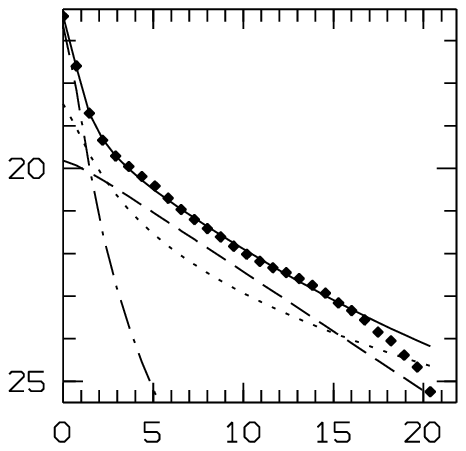}}
\put(5.9,3.0){%
\includegraphics[bb = 87 87 209 209,height=\proftab,clip]{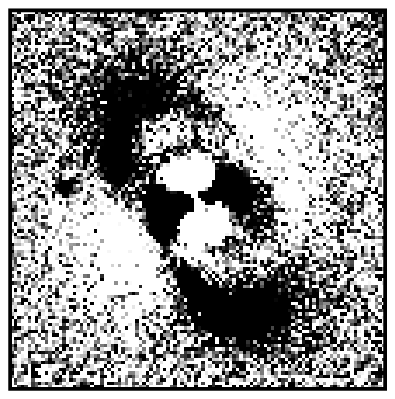}}
\put(0.0,00.3){\rotatebox{90}{HE\,1348--1758}}
\put(0.3,00){%
\includegraphics[bb = 91 642 244 771,height=\proftab,clip]{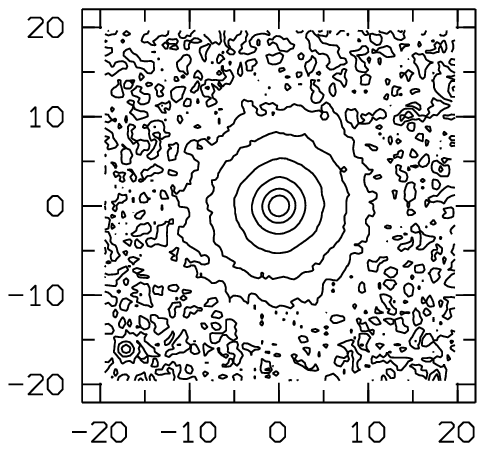}}
\put(3.1,00){%
\includegraphics[bb = 91 642 244 771,height=\proftab,clip]{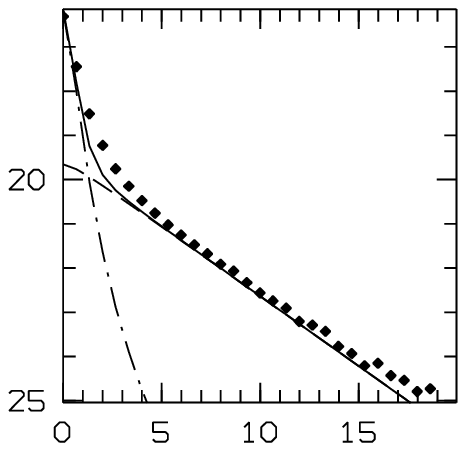}}
\put(5.9,00){%
\includegraphics[bb = 88 88 200 200,height=\proftab,clip]{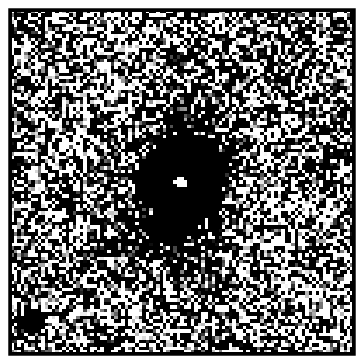}}
\end{picture}
  \caption{Two examples of unsatisfactory fits, requiring user intervention. 
    For HE 1017--0305 a excessively large spheroid is fit due to
    surplus in surface brightness at 6\arcsec to 9\arcsec. For HE
    1348--1758 we forced an unresolved nuclear source instead of a
    spheroid. Notation is as in Fig~\protect{\ref{fig:datamed}}}
  \label{fig:badfits}
\end{figure}

\subsection{Results}  \label{sec:res}

For all of our objects we were able to obtain satisfactory fits. Some
examples of objects, profiles, and residual images after subtracting
the models are presented in Figs~\ref{fig:datamed} and~\ref{fig:datalow}.

In some of the residual images, little or no structure was left at the
location of the galaxies. This is the case for HE\,1348$-$1758 and
HE\,0914$-$0031.  In the majority of the objects, strong features were
present, mainly indicating the limitations of the azimuthally
symmetric models.  These morphological features will also influence
the fit itself and may render it less reliable. An example is
HE\,$1017-0305$, where a pair of counter-coiled spiral arms, or tidal
features, causes a significant surplus of flux at larger radii which
mimics the contribution of an unphysically large spheroidal component.
The resulting fit is shown in Fig~\ref{fig:badfits}.  A better fit is
obtained when the fitting area is restricted to regions unaffected by
the extended arms (Fig\ref{fig:datalow}, top line).

The magnitudes of the components were computed from the best-fitting
model parameters, which already contain the total fluxes for each
component. Other methods are possible such as using the obtained
nuclear model to subtract the nuclear source and then determining the
galaxy flux from growth curves of the remaining host galaxy. This
method yields a more precise nuclear-to-ratio as it is strictly
flux-conserving. For the goal of computing the fluxes of several
galaxy components separately, this is not as straightforward. For our
data we tested and compared both methods and found that they agree
extremely well. Fluxes taken from the model parameters are, on
average, brighter by 2 per cent, with an rms scatter of $\pm$ 4 per
cent. 

\begin{figure}
  \includegraphics[bb = 45 58 317 258,clip,width=\colwidth]{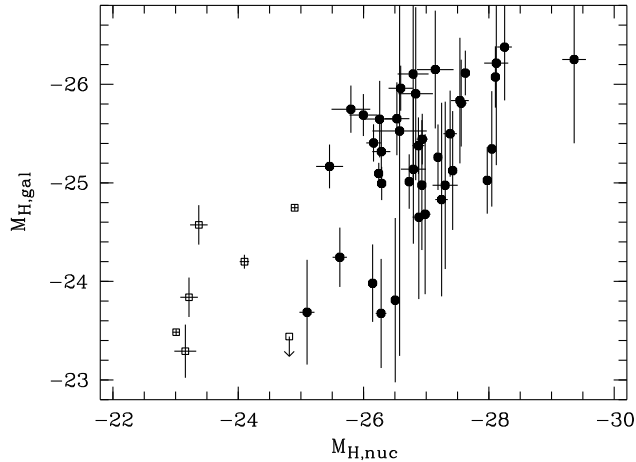}
  \caption{Nuclear versus spheroid luminosities. The circles are from
    the medium redshift, squares from the low redshfit sample.
    Errorbars are based on  dedicated simulations. The arrow
    indicates an upper limit for the speroid.
  \label{fig:b2nuc}}
\end{figure}

Magnitude errors were estimated using dedicated simulations (described
in the section~\ref{sec:sim}) by interpolating the uncertainties of
the two most similar simulated objects. We did not include error
estimates concerning the systematic differences between fitted and
observed profile but included the uncertainties raised by the errors
of PSF radius and the sky background level. In Fig~\ref{fig:b2nuc}
we plot magnitudes of the host galaxy and the nuclear component
along with their uncertainties.

\section{Conclusions}

We have presented a versatile method to describe the light
distribution of quasar host galaxy images. Particular attention was
paid to a careful quantitative modelling of the PSF underlying the
quasar images, in order to make the decomposition of the quasar into
host galaxy and nuclear components as reliable as possible, even using
non-optimal observing material.

The simulations shown in the present paper show that our method is
well suited to derive accurate and unbiased host galaxy luminosities
with realistic error estimates.

We demonstrated the usefulness of the method for two small samples of
local Seyfert galaxies and medium redshift quasars. The latter was
observed with the main intention to constrain the host galaxy
luminosity distribution function.  Preliminary results were presented
by \citet{wiso01} a more detailed account is given by \citet{kuhl03}.

We have already applied the algorithm to other data.  For a sample of
$\sim 20$ low-redshift quasars ($z<0.2$), which we observed in several
optical and near-infrared wavebands, we have derived spectral energy
distributions and stellar population descriptions \citep{jahn03}.
Additional applications can easily be conceived.

\section{Acknowledgements}
This work was supported by the DFG under grants Wi~1369/5--1 and
Re~353/45-3. We made use of observing time at ESO with the programme
61.B-0300. KJ acknowledges support by the `Studienstiftung des
Deutschen Volkes'.

\bibliographystyle{mn2e}
\bibliography{bkbib}

\begin{thebibliography}{}

\bibitem[\protect\citeauthoryear{{Abraham}, {Crawford} \& {McHardy}}{{Abraham}
  et~al.}{1992}]{abra92}
{Abraham} R.~G.,  {Crawford} C.~S.,    {McHardy} I.~M.,  1992, ApJ, 401, 474

\bibitem[\protect\citeauthoryear{de Vaucouleurs}{de~Vaucouleurs}{1948}]{deva48}
de Vaucouleurs G.,  1948, Ann. Astrophys., 11, 247

\bibitem[\protect\citeauthoryear{Duda \& Hart}{Duda \& Hart}{1973}]{duda73}
Duda R.,  Hart P.,  1973, Pattern Classification and Scene Analysis.
Wisley

\bibitem[\protect\citeauthoryear{{Falomo}, {Kotilainen} \& {Treves}}{{Falomo}
  et~al.}{2001}]{falo01}
{Falomo} R.,  {Kotilainen} J.,    {Treves} A.,  2001, ApJ, 547, 124

\bibitem[\protect\citeauthoryear{Freeman}{Freeman}{1970}]{free70}
Freeman K.~C.,  1970, ApJ, 160, 812

\bibitem[\protect\citeauthoryear{{Jahnke}, {Kuhlbrodt} \& {Wisotzki}}{{Jahnke}
  et~al.}{2003}]{jahn03}
{Jahnke} K.,  {Kuhlbrodt} B.,    {Wisotzki} L.,  2003, astro-ph/0311123

\bibitem[\protect\citeauthoryear{Kuhlbrodt}{Kuhlbrodt}{2003}]{kuhl03}
Kuhlbrodt B.,  2003, PhD thesis, Universit\"at Hamburg

\bibitem[\protect\citeauthoryear{Kukula, Dunlop, McLure, Miller, Percival, Baum
  \& O'Dea}{Kukula et~al.}{2001}]{kuku01}
Kukula M.~J.,  Dunlop J.~S.,  McLure R.~J.,  Miller L.,  Percival W.,  Baum
  S.~A.,    O'Dea C.~P.,  2001, MNRAS, 326, 1533

\bibitem[\protect\citeauthoryear{{Lehnert}, {van Breugel}, {Heckman} \&
  {Miley}}{{Lehnert} et~al.}{1999}]{lehn99}
{Lehnert} M.~D.,  {van Breugel} W.~J.~M.,  {Heckman} T.~M.,    {Miley} G.~K.,
  1999, ApJS, 124, 11

\bibitem[\protect\citeauthoryear{McLeod \& McLeod}{McLeod \&
  McLeod}{2001}]{mcle01}
McLeod K.~K.,  McLeod B.~A.,  2001, ApJ, 546, 782

\bibitem[\protect\citeauthoryear{{McLure}, {Dunlop} \& {Kukula}}{{McLure}
  et~al.}{2000}]{mclu00}
{McLure} R.~J.,  {Dunlop} J.~S.,    {Kukula} M.~J.,  2000, MNRAS, 318, 693

\bibitem[\protect\citeauthoryear{{McLure}, {Kukula}, {Dunlop}, {Baum}, {O'Dea}
  \& {Hughes}}{{McLure} et~al.}{1999}]{mclu99}
{McLure} R.~J.,  {Kukula} M.~J.,  {Dunlop} J.~S.,  {Baum} S.~A.,  {O'Dea}
  C.~P.,    {Hughes} D.~H.,  1999, MNRAS, 308, 377

\bibitem[\protect\citeauthoryear{Moffat}{Moffat}{1969}]{moff69}
Moffat A. F.~J.,  1969, A\&A, 3, 455

\bibitem[\protect\citeauthoryear{Press, Teukolsky, Vetterling \&
  Flannery}{Press et~al.}{1995}]{pres95}
Press W.~H.,  Teukolsky S.~A.,  Vetterling W.~T.,    Flannery B.~P.,  1995,
  Numerical recipes in C, 2nd edn.
Cambridge University Press

\bibitem[\protect\citeauthoryear{{Ravindranath}, {Ho}, {Peng}, {Filippenko} \&
  {Sargent}}{{Ravindranath} et~al.}{2001}]{ravi01}
{Ravindranath} S.,  {Ho} L.~C.,  {Peng} C.~Y.,  {Filippenko} A.~V.,
  {Sargent} W.~L.~W.,  2001, AJ, 122, 653

\bibitem[\protect\citeauthoryear{{Ridgway}, {Heckman}, {Calzetti} \&
  {Lehnert}}{{Ridgway} et~al.}{2001}]{ridg01}
{Ridgway} S.~E.,  {Heckman} T.~M.,  {Calzetti} D.,    {Lehnert} M.,  2001, ApJ,
  550, 122

\bibitem[\protect\citeauthoryear{{Schade}, {Lilly}, {Le Fevre}, {Hammer} \&
  {Crampton}}{{Schade} et~al.}{1996}]{scha96}
{Schade} D.,  {Lilly} S.~J.,  {Le Fevre} O.,  {Hammer} F.,    {Crampton} D.,
  1996, ApJ, 464, 79

\bibitem[\protect\citeauthoryear{Stetson}{Stetson}{1987}]{stet87}
Stetson P.~B.,  1987, PASP, 99, 191

\bibitem[\protect\citeauthoryear{{Wadadekar}, {Robbason} \&
  {Kembhavi}}{{Wadadekar} et~al.}{1999}]{wada99}
{Wadadekar} Y.,  {Robbason} B.,    {Kembhavi} A.,  1999, AJ, 117, 1219

\bibitem[\protect\citeauthoryear{Wisotzki, Christlieb, Bade, Beckmann,
  K\"ohler, Vanelle \& Reimers}{Wisotzki et~al.}{2000}]{wiso00}
Wisotzki L.,  Christlieb N.,  Bade N.,  Beckmann V.,  K\"ohler T.,  Vanelle C.,
     Reimers D.,  2000, A\&A, 358, 77

\bibitem[\protect\citeauthoryear{{Wisotzki}, {Kuhlbrodt} \&
  {Jahnke}}{{Wisotzki} et~al.}{2001}]{wiso01}
{Wisotzki} L.,  {Kuhlbrodt} B.,    {Jahnke} K.,  2001, in M\'arquez I.,  et~al.
  eds, {\em QSO Hosts and Their Environments} Kluwer

\end{thebibliography}
\end{document}